\documentclass[conference]{IEEEtran}

\usepackage{courier}

\usepackage{fancyhdr}
\usepackage{hyperref}
\usepackage{color}

\renewcommand{\thispagestyle}[1]{} % do nothing

\renewcommand{\headrulewidth}{0pt}
\renewcommand{\footrulewidth}{0.4pt}

\ifCLASSINFOpdf
\usepackage[pdftex]{graphicx}
\graphicspath{{/home/isro1/Documents/respond/thesis/conf_papers}}
\DeclareGraphicsExtensions{.pdf,.jpeg,.png}
\else
\usepackage[dvips]{graphicx}
\DeclareGraphicsExtensions{.eps}
\fi
\usepackage[cmex10]{amsmath}
\usepackage{array}
\usepackage[caption=false,font=footnotesize]{subfig}
\begin{document}

\title{Performance Comparison of Linear Prediction based Vocoders in Linux Platform}

%\author{\IEEEauthorblockN{Lani Rachel Mathew, Ancy S. Anselam and Sakuntala S. Pillai, \emph{Senior Member IEEE}}
%\author{\IEEEauthorblockN{Lani Rachel Mathew, Ancy S. Anselam and Sakuntala S. Pillai}
%\IEEEauthorblockA{Department of Electronics and Communication Engineering\\
%Mar Baselios College of Engineering and Technology, Nalanchira\\
%Thiruvananthapuram 695 015, Kerala, India\\
%lanirachel@gmail.com, ancy\_anselam@yahoo.co.in, sakuntala.pillai@gmail.com } %\and
%\IEEEauthorblockN{Sakuntala S. Pillai}
%\IEEEauthorblockA{\emph{Senior Member, IEEE}\\
%Mar Baselios College of Engg. \& Technology\\
%Thiruvananthapuram, India\\
%Email: sakuntala.pillai@gmail.com}


\author{\IEEEauthorblockN{Lani Rachel Mathew\IEEEauthorrefmark{1},
Ancy S. Anselam\IEEEauthorrefmark{2} and
Sakuntala S. Pillai\IEEEauthorrefmark{3}
\\ Department of Electronics and Communication Engineering\\
Mar Baselios College of Engineering and Technology, Nalanchira\\
Thiruvananthapuram 695 015, Kerala, India}
\IEEEauthorblockA{\IEEEauthorrefmark{1}M.Tech Student, \texttt{lanirachel@gmail.com}}%\\Department of Electronics and Communication Engineering\\
%Mar Baselios College of Engineering and Technology, Nalanchira\\
%Thiruvananthapuram 695 015, Kerala, India} %Email: see http://www.michaelshell.org/contact.html}
\IEEEauthorblockA{\IEEEauthorrefmark{2}Assistant Professor, \texttt{ancy\_anselam@yahoo.co.in}}%, Department of Electronics and Communication Engineering\\
%Mar Baselios College of Engineering and Technology, Nalanchira\\
%Thiruvananthapuram 695 015, Kerala, India}
\IEEEauthorblockA{\IEEEauthorrefmark{3}Dean (R~\&~D), \texttt{sakuntala.pillai@gmail.com}}}
%Telephone: (800) 555--1212, Fax: (888) 555--1212}


\pagestyle{fancy}
\chead{\textit{\textbf{\textcolor{red}{International Journal of Engineering Trends and Technology (IJETT) – Volume 10 Issue 11- April 2014}}}}
%\chead{\textit{\textbf{\textcolor{red}{International Journal of Engineering Trends and Technology (IJETT)}}}}
\cfoot{\url{http://www.ijettjournal.org}}
\lfoot{ISSN: 2231-5381}
\rfoot{Page \thepage}
\renewcommand{\headrulewidth}{0pt}
\renewcommand{\footrulewidth}{0.4pt}

\maketitle

\begin{abstract}
%\boldmath
Linear predictive coders form an important class of speech coders. This paper describes the software level implementation of linear prediction based vocoders, viz. Code Excited Linear Prediction (CELP), Low-Delay CELP (LD-CELP) and Mixed Excitation Linear Prediction (MELP) at bit rates of 4.8 kb/s, 16 kb/s and 2.4 kb/s respectively. The C programs of the vocoders have been compiled and executed in Linux platform. Subjective testing with the help of Mean Opinion Score test has been performed. Waveform analysis has been done using Praat and Adobe Audition software. The results show that MELP and CELP produce comparable quality while the quality of LD-CELP coder is much higher, at the expense of higher bit rate.
\newline \newline
\emph{Keywords - Vocoder, linear prediction, code excited, low delay, mixed excitation, CELP, LD-CELP, MELP, Praat, Linux}

\end{abstract}


% \fi
\IEEEpeerreviewmaketitle

\section{Introduction}
Speech coding is the encoding of speech signals to enable transmission at bit rates lower than that of the original digitized speech. The human auditory system can capture only certain aspects of a speech signal. Thus, perceptually relevant information of a speech signal can be extracted to produce an equivalent-sounding wave at a much lower bandwidth. \newline \indent Linear prediction is a widely used compression technique in which past samples of a signal are stored and used to predict the next sample \cite{Makh}. In the basic linear predictive coder prototype, prediction is done over a time interval of one pitch period using adaptive linear delay and gain factors. This basic prototype produces intelligible but artificial-sounding speech output, and various techniques have been researched to improve the perceptual quality. \newline \indent Variants to the linear prediction coders are Code Excited Linear Prediction (CELP) and Low-Delay CELP (LD-CELP) which use forward and backward linear prediction respectively, along with the \emph{codebooks}, i.e. lookup tables with codevectors corresponding to speech residual signals \cite{atal,Chen}. In Mixed Excitation Linear Prediction (MELP) \cite{alan}, an additional classification of speech is introduced - the jittery voiced speech. Mixed excitation, i.e the mixing of periodic and noise excitation, is another distinguishing feature of the MELP model. \newline \indent This paper aims at comparing the three types of linear prediction based vocoders in terms of their bit rate and perceptual quality. In comparing vocoders, subjective testing of the voice quality is a major step in the evaluation of a vocoder. A vocoder will finally be accepted in the market only if humans are satisfied with the voice quality. Keeping this fact in consideration, subjective evaluation using the Mean Opinion Score (MOS) test has been conducted. \newline \indent The paper is organized as follows: Section II gives a brief overview of the vocoder algorithms. The bit allocation and bit rate calculations are also described. Section III describes the method adopted in implementing and testing the vocoder. The results obtained and their implications are discussed in Section IV, followed by concluding remarks.
\begin{figure}[!t]
\centering
\includegraphics[width=2.5in]{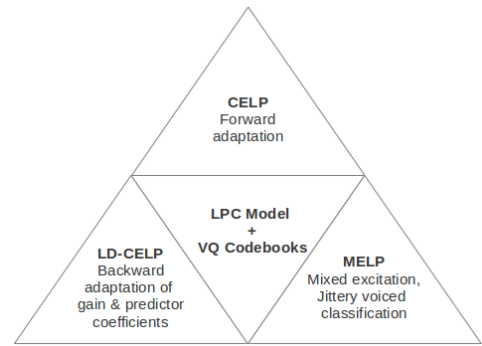}
\caption{LPC model as the core of CELP, LD-CELP and MELP algorithms}
\label{fig:core}
\end{figure}

\begin{figure*}[!t]
\centerline{\subfloat[Encoder]{\includegraphics[scale=0.5]{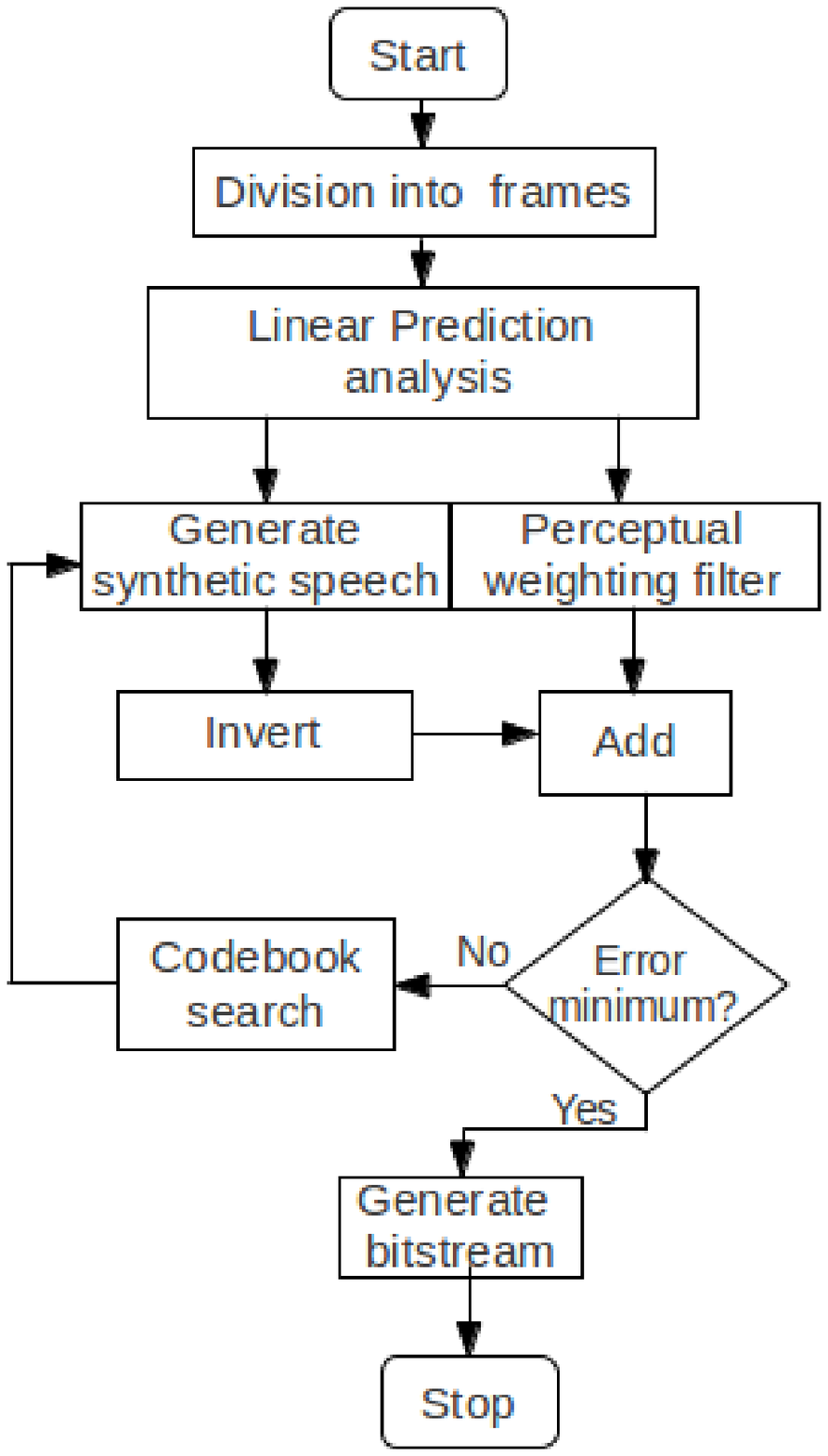}%
\label{fig:encoder}}
\hfil
\subfloat[Decoder]{\includegraphics[scale=0.5]{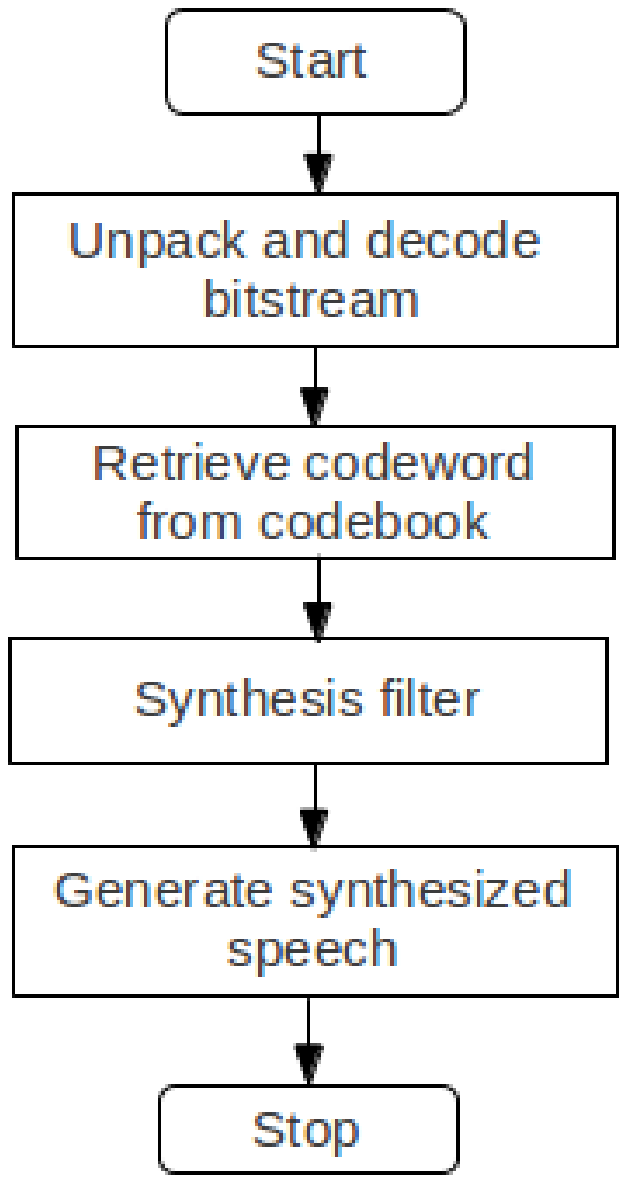}%
\label{fig:decoder}}}
\caption{Generalized Design Flow for Linear Prediction based Vocoder}
\label{fig:designflow}
\end{figure*}

%-----------------------------------------------------------------------------------------------------------------------

%\section{Linear Prediction Coding: An Overview}

%-----------------------------------------------------------------------------------------------------------------------
\section{Vocoder Algorithms}

The Linear Prediction model forms the core of the CELP, LD-CELP and MELP algorithms, as depicted in Fig.~\ref{fig:core}. In the LPC model, speech signals are classified into voiced and unvoiced signals. Voiced signals are generated when the vocal cords vibrate and are represented in the LPC source-filter model as periodic excitation. Unvoiced signals are generated by turbulence of air in the vocal tract and do not involve the vocal cords. These signals are usually represented as white Gaussian noise. 
\newline \indent The LPC coder consists of a linear predictor having adaptive delay and gain factors \cite{Makh}. Since there are sounds in speech that are produced by a combination of voiced and unvoiced signals, it has been observed that important perceptual speech information gets missed out from the predictor output. Also a slight error in the predictor coefficients will lead to more speech information being missed out. This unaccounted residual output of the LPC coder contains important data on how the sound signal is perceived by the human auditory system. 

\subsection{Code Excited Linear Prediction Algorithm}
It was proposed in \cite{atal} that the prediction residual signal could be used to enhance the perceptual quality of the coder output. A codebook containing a list of codevectors is searched to obtain a closest match to the residual signal. The index of the codevector is selected such that minimization of the perceptually weighted error metric is obtained. The codec earned its name due to the use of codebooks to obtain \emph{codes} for modeling the speech signal. The same codebook is available at both the transmitter and the receiver. At the receiver, the index is used to obtain the codevector and use it in the synthesis filters. 
\newline \indent  CELP uses the Analysis-by-Synthesis (AbS) method in which the transmitter \emph{analyzes} the signal to produce linear prediction coefficients, and then uses these coefficients to \emph{synthesize} the speech signal within the transmitter itself. An error signal is generated and codevectors are selected from the codebook in order to minimize the perceptually weighted mean square error. 
\newline \indent  In the CELP Transmitter, the transmitter first splits the input speech into frames of around 30 ms. Short-term linear prediction is performed, i.e. formants (peaks of the spectral envelope) are estimated for each input frame. The transfer function of the perceptual weighting filter of CELP is given by the following equation.
\begin{equation}
  W(\emph{z})= \frac{1-Q(\emph{z})}{1-Q(\frac{z}{\gamma})}
  \label{eq:W(z)}
  \end{equation}
where

\begin{equation}
  Q(\emph{z})= \sum\limits_{i=1}^M q_{i}z^{-i}
\label{eq:Q(z)}
    \end{equation}

\begin{equation}
Q(\frac{z}{\gamma})=\sum\limits_{i=1}^M \gamma^{i} q_{i}z^{-i}, 0 < \gamma < 1
  \label{eq:Q(z/gamma)}
  \end{equation}

where M is the LPC predictor order and \(q_{i}\)'s are the quantized LPC coefficients. After finding the short term LPC coefficients, each frame is split into four subframes, i.e. 7.5 ms each, which are given as input to the long-term prediction filter. The pitch and the intensity of the speech signal are estimated. An optional post filtering stage may be added after decoding to enhance the quality of the output signal. The CELP bit allocation \cite{wai} is shown below:
\begin{center}
\begin{tabular}{|c|c|c|}
\hline
\textbf{Parameter} & \textbf{No./frame} & \textbf{Total bits/frame}\\
\hline

	Linear prediction coefficients   & 10  & 34\\ 
	Pitch period   & 4  & 28\\
	Adaptive codebook gain & 4 & 20\\
	Stochastic codebook index & 4 & 36\\
	Stochastic codebook gain & 4 & 20\\
	Synchronization & 1 & 1\\
	Error correction & 4 & 4\\
	Future expansion & 1 & 1\\\hline
	Total bits/30 ms frame &   &   144\\ \hline

\end{tabular}
\label{tab:bitalloc_celp}
\end{center}

The 30 ms frame of CELP corresponds to 240 samples for a sampling rate of 8000 Hz. Thus the bit rate of CELP is 144/30ms = 4.8 kb/s.
\begin{figure*}[!t]
\subfloat[Original Sound File]{\includegraphics[scale=0.45]{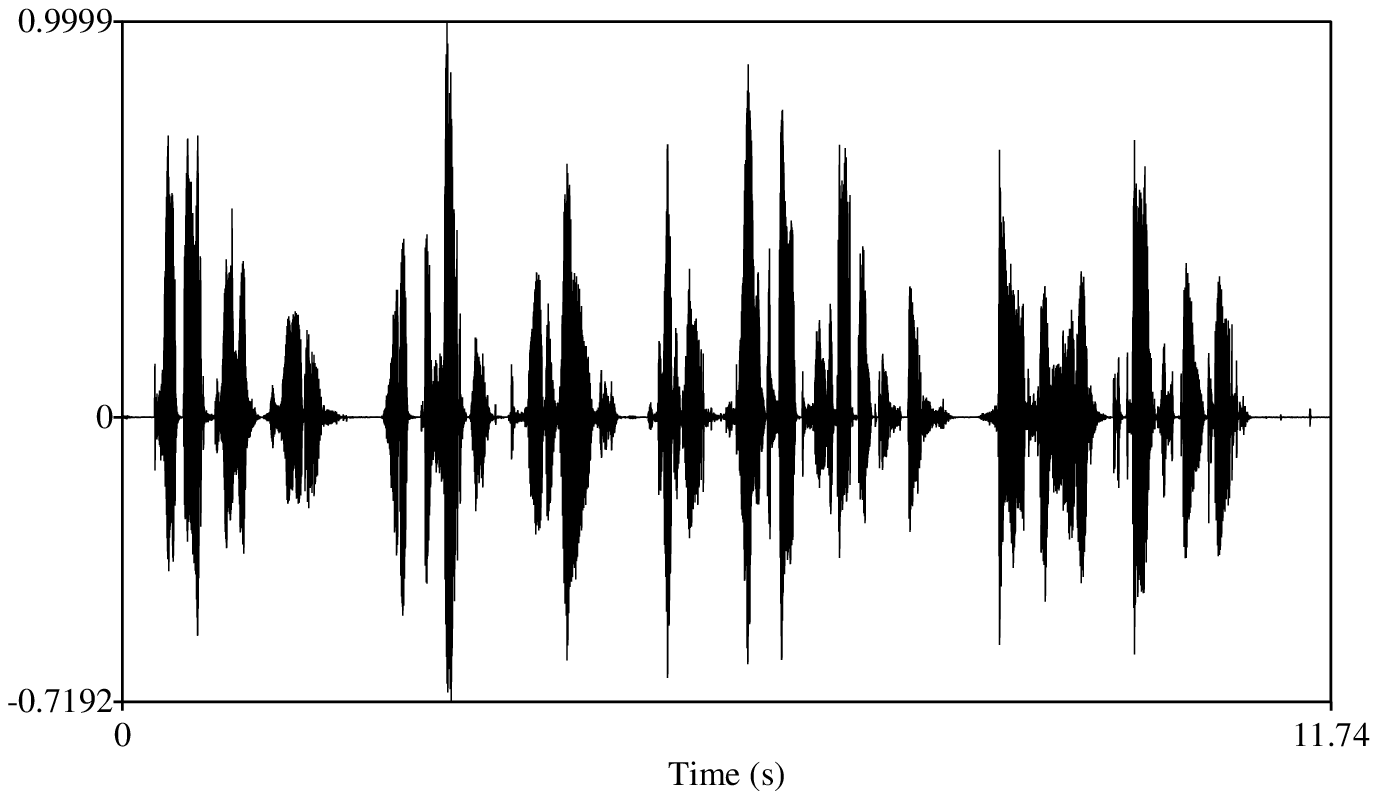}%
\label{fig:time_orig}}
\hfil
\subfloat[CELP synthesized output]{\includegraphics[scale=0.45]{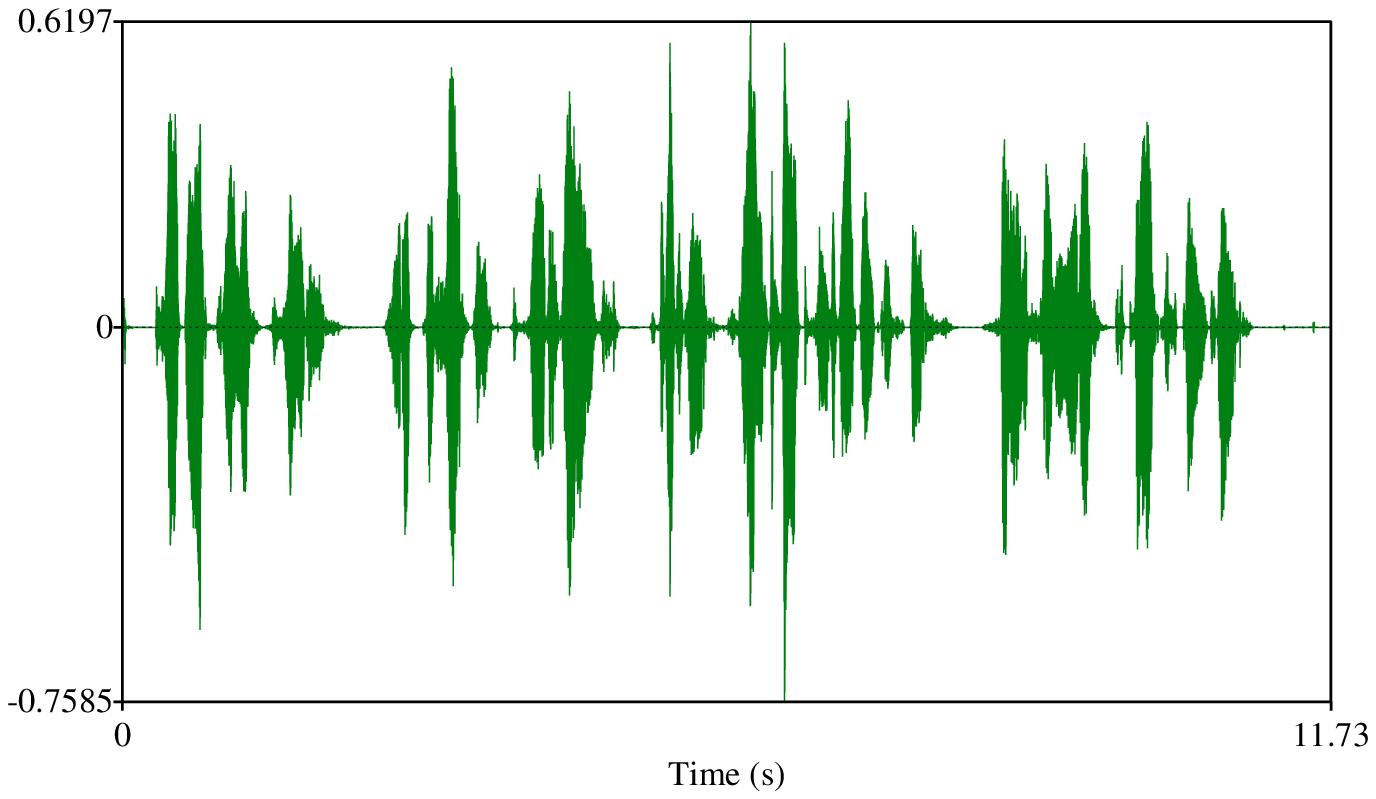}%
\label{fig:time_celp}}
\newline
\subfloat[LD-CELP synthesized output]{\includegraphics[scale=0.45]{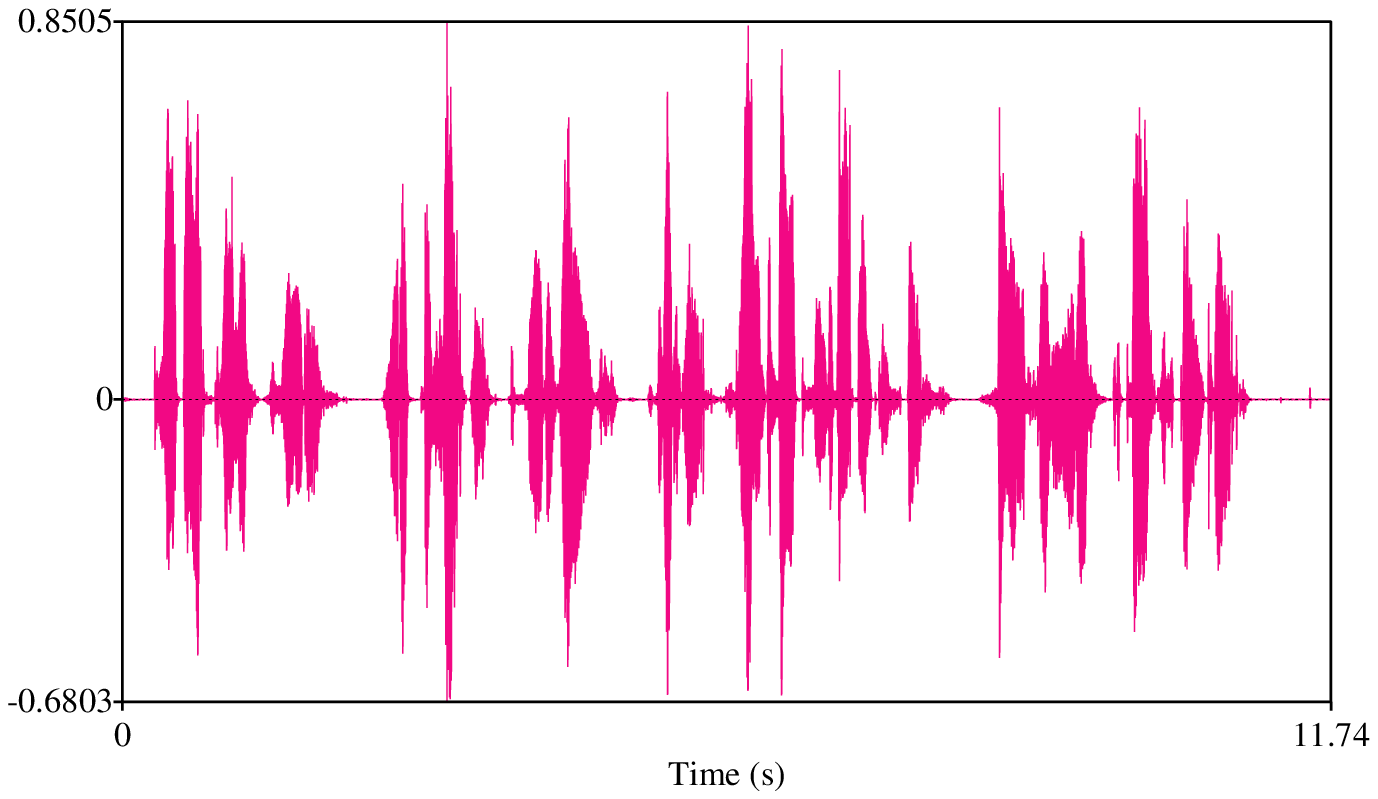}%
\label{fig:time_ldcelp}}
\hfil
\subfloat[MELP synthesized output]{\includegraphics[scale=0.45]{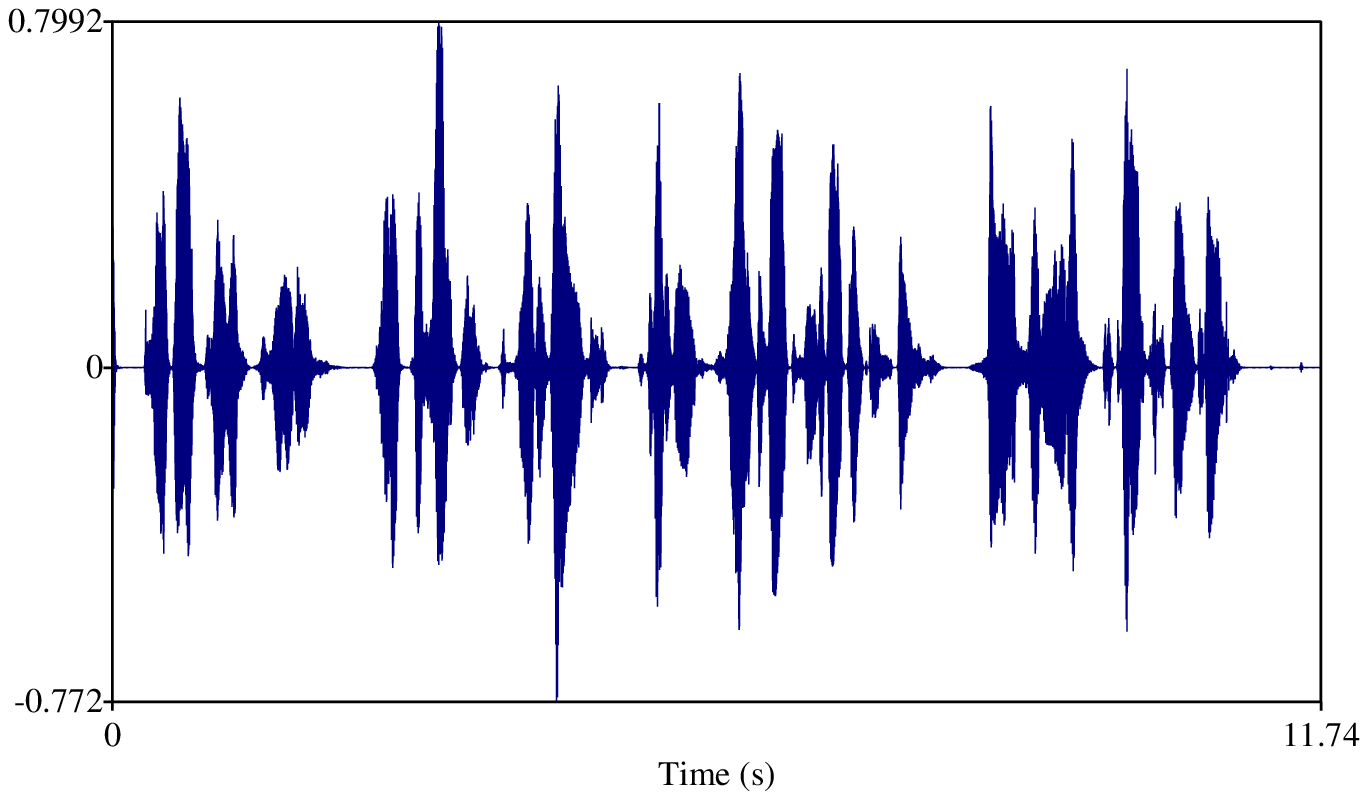}%
\label{fig:time_melp}}
\caption{Time domain representation }
\label{fig:time}
\end{figure*}

%-----------------------------------------------------------------------------------------------------------------------
\subsection{Low Delay Code Excited Linear Prediction Algorithm}
The CELP and LD-CELP algorithms differ only in the type of adaptation - forward and backward respectively - in which linear prediction is performed. Low delay is achieved by the use of a backward-adaptive predictor and short excitation vectors (5 samples each) \cite{Chen}. Only the index of the excitation codebook is transmitted - all other parameters are updated by backward adaptation of previously quantized speech. LD-CELP uses a modified system function for the weighting filter as given below.
\begin{equation}
  W(\emph{z})= \frac{1-Q(\frac{z}{\gamma_{1}})}{1-Q(\frac{z}{\gamma_{2}})}, 0 < \gamma_{2} <\gamma_{1} \le{1}
  \label{eq:ldcelpW(z)}
  \end{equation}
where the parameters \(\gamma_{1}\) and \(\gamma_{2}\) are tuned to optimize the quality of the coded speech and Q(z) is given by the expression given below.

The LD-CELP bit allocation is shown below \cite{voip}:\newline
\begin{center}
\begin{tabular}
{|c|c|}\hline
\textbf{Parameter} & \textbf{No. of Bits} \\
\hline
	Excitation Index  & 7  \\ 
	Excitation Gain   & 2 \\
	Sign codebook gain & 1\\ \hline
	Total bits/2.5 ms vector &   10\\ 
	Total bits/10 ms frame &   40\\ \hline

\end{tabular}
\label{tab:bitalloc_ldcelp}
\end{center}
5 samples, i.e. a vector corresponds to 0.625 ms for a sampling rate of 8000 Hz. Thus the bit rate of LD-CELP is 10/0.625ms = 16 kb/s.

%-----------------------------------------------------------------------------------------------------------------------
\subsection{Mixed Excitation Linear Prediction Algorithm}
In the MELP coder, there are three classifications for the speech signal - voiced, unvoiced and jittery voiced. The third classification is done when voicing transitions occur, i.e. when aperiodic but not completely random excitations occur. Another feature is that the shape of the excitation pulse is also extracted from the input signal. Pulse shaping filters and noise shaping filters are used to filter the pulse train and white noise excitations. `Mixed excitation' refers to the total excitation signal which is the sum of the filtered output periodic and noise excitations. 
The MELP bit allocation is described in the table below \cite{supplee}:
\begin{center}
\begin{tabular}{|c|c|c|} 
\hline
\textbf{Parameter} & \textbf{Voiced} & \textbf{Unvoiced}\\
\hline
	LSF parameters  & 25 & 25 \\ 
	Fourier magnitudes & 8 & - \\ 
	Gain (2 per frame) & 8 & 8 \\ 
	Pitch, overall voicing  & 7 & 7 \\ 
	Bandpass voicing  & 4 & - \\ 
	Aperiodic flag   & 1 & - \\
	Error protection   & - & 13 \\
	Sync bit   & 1 & 1 \\ \hline
	Total bits/22.5 ms frame & 54 & 54\\ \hline
\end{tabular}
\label{tab:bitalloc_melp}
\end{center}

The 22.5 ms frame of MELP corresponds to 180 samples for a sampling rate of 8000 Hz. Thus the bit rate of MELP is 54/22.5ms = 2.4 kb/s.

\begin{figure*}[!t]
%\centerline{\subfloat[Original spectrogram]{\includegraphics[scale=0.15]{spec_orig}%
\subfloat[Original Spectrogram]{\includegraphics[scale=0.15]{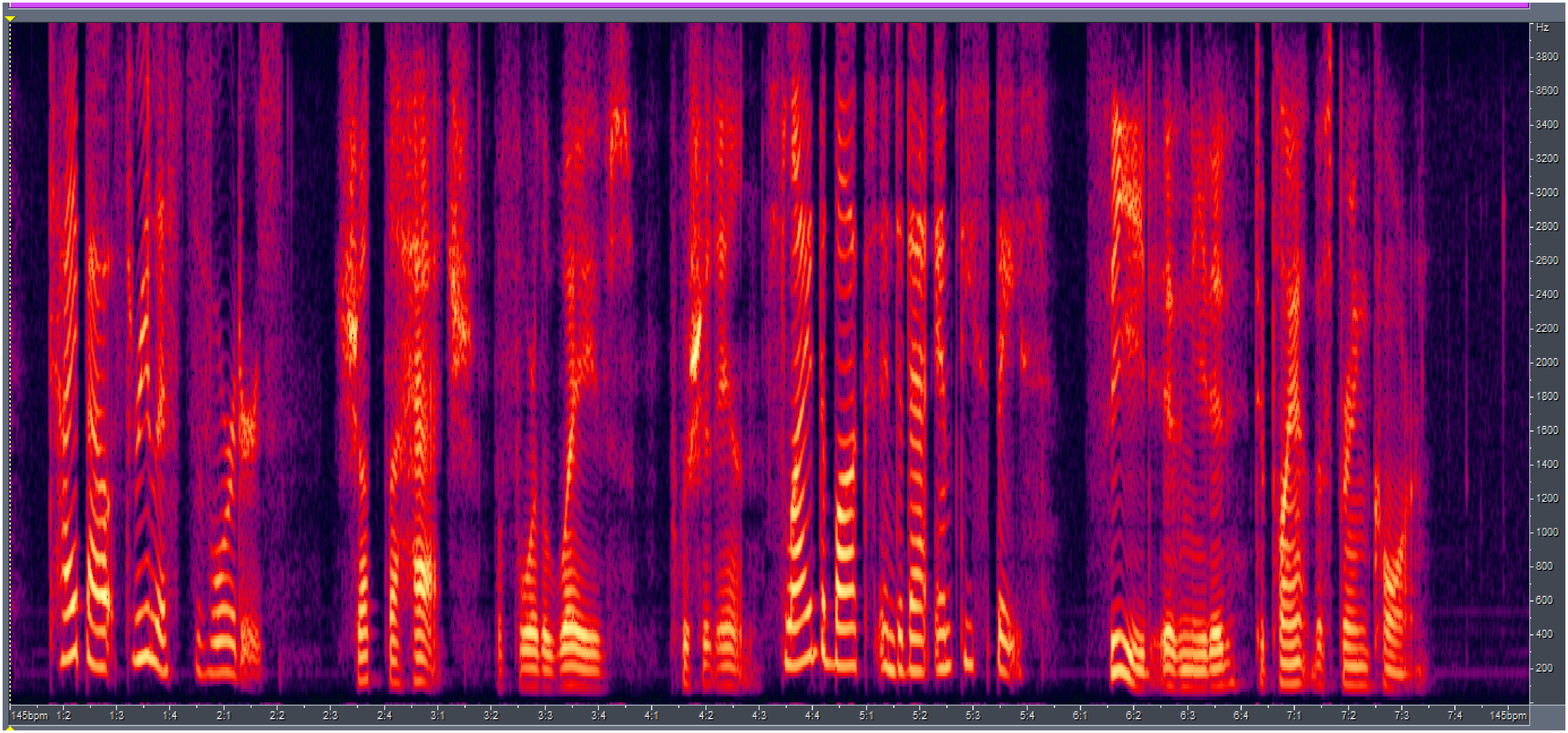}%
\label{fig:spec_orig}}
\hfil
\subfloat[CELP Spectrogram]{\includegraphics[scale=0.15]{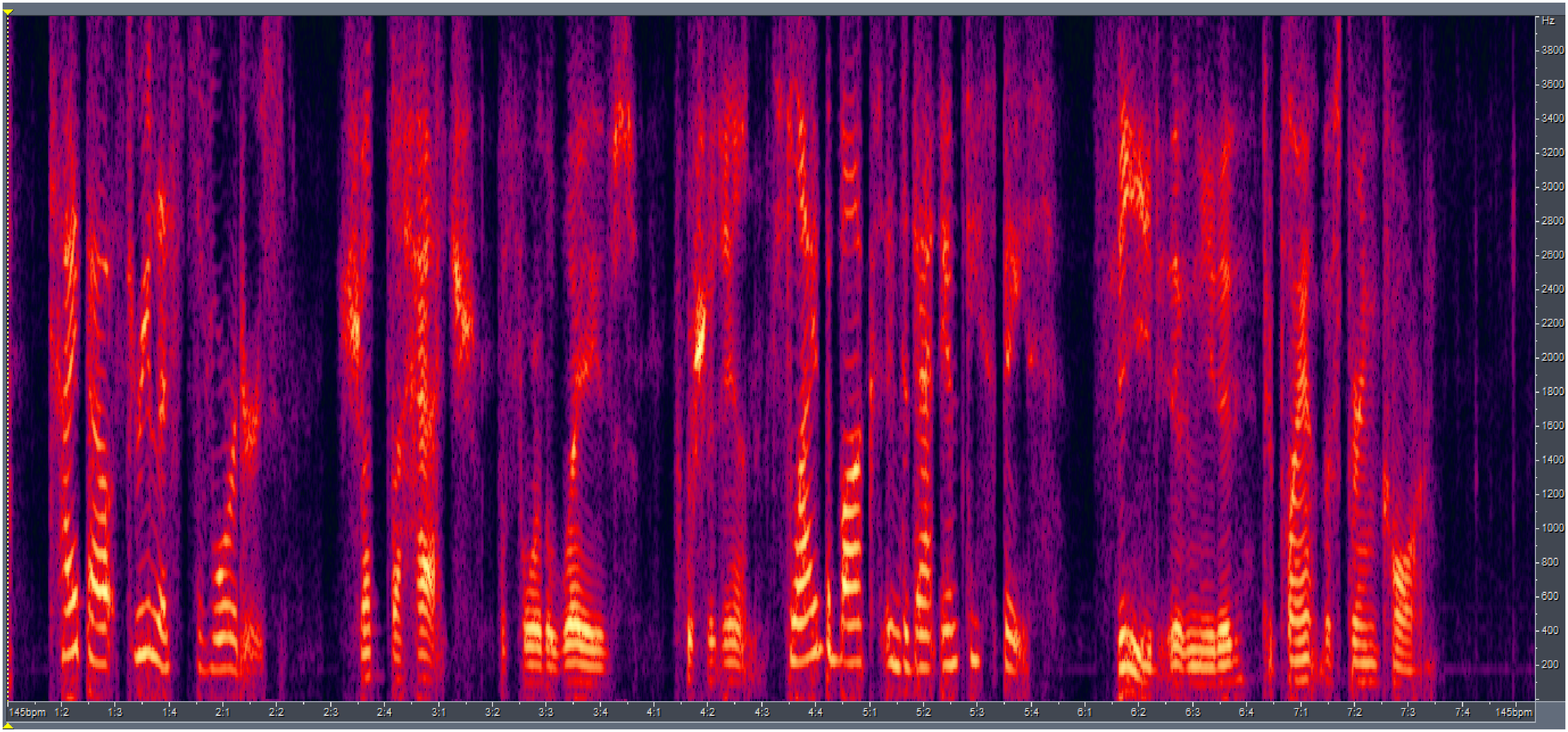}%
\label{fig:spec_celp}}
\newline
\subfloat[LD-CELP Spectrogram]{\includegraphics[scale=0.15]{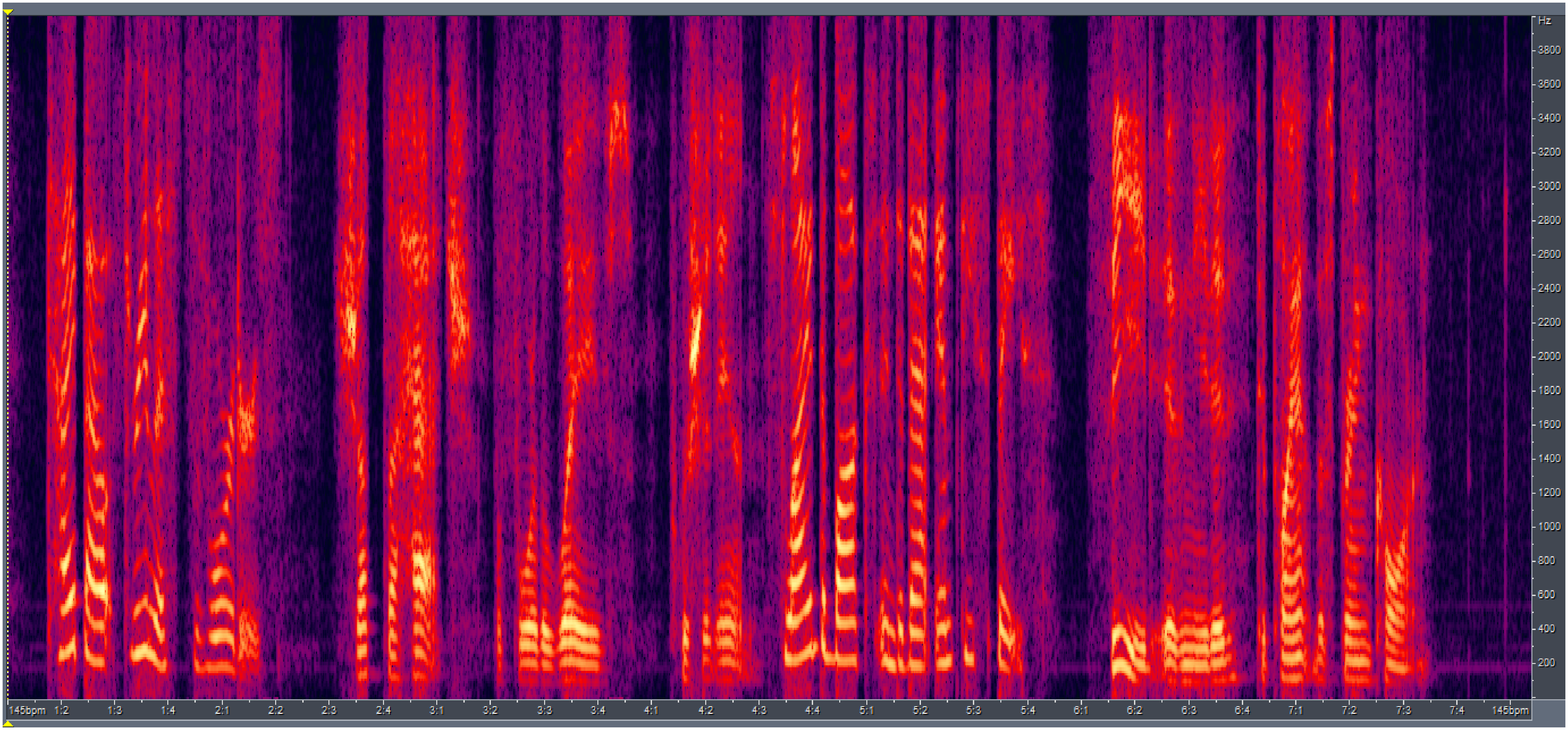}%
\label{fig:spec_ldcelp}}
\hfil
\subfloat[MELP Spectrogram]{\includegraphics[scale=0.15]{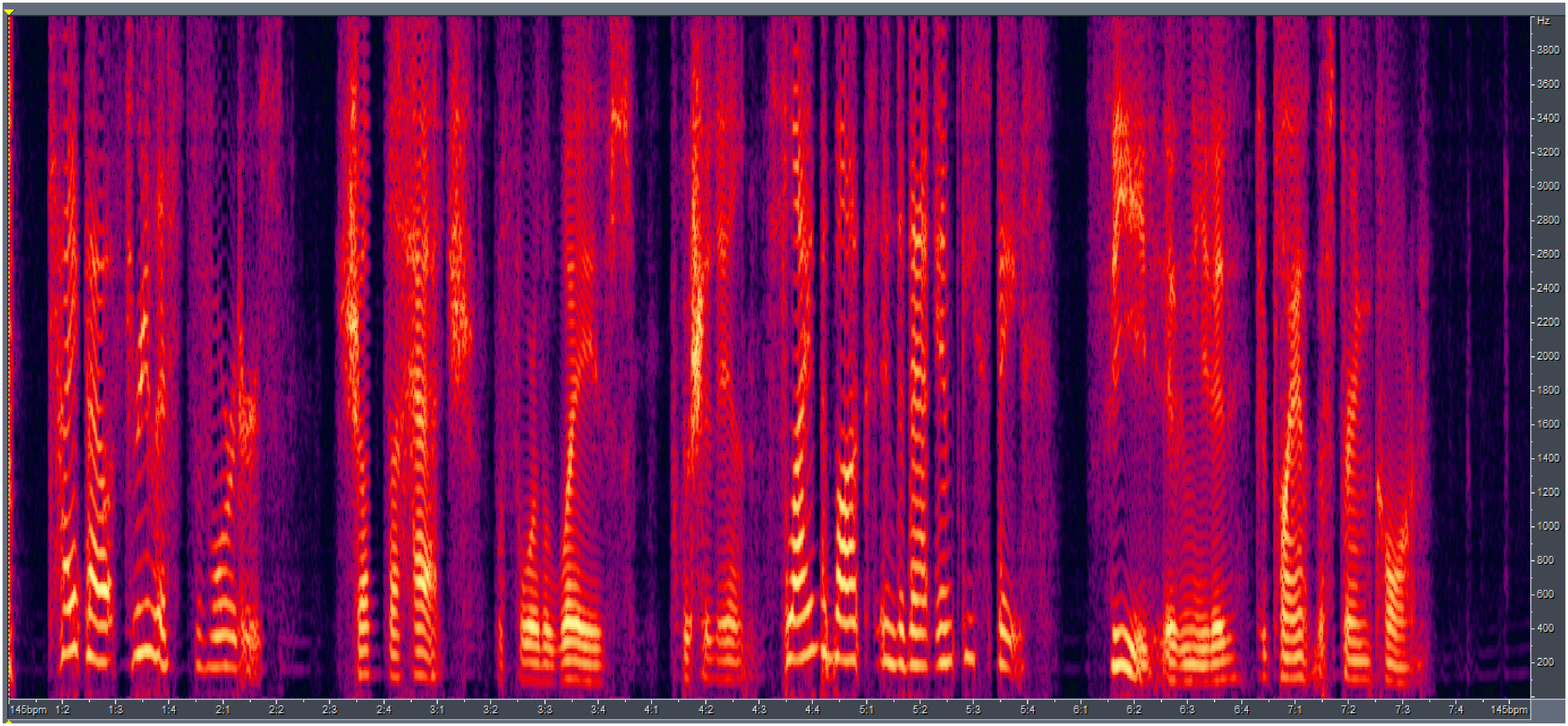}%
\label{fig:spec_melp}}
\caption{Spectrograms of speech signals}
\label{fig:spectrogram}
\end{figure*}

%-----------------------------------------------------------------------------------------------------------------------
\section{Method}
The C programs of the vocoders were compiled with GCC(GNU Compiler Collection) and built using the GNU Make utility in a Linux platform. For waveform conversions, Sound eXchange (SoX) software was used. SoX is an open-source tool for speech file manipulations. The analysis and synthesis commands used in each vocoder are enlisted below:

\begin{enumerate}
  \item CELP commands 
  	\begin{enumerate}
    	\item \emph{Analysis}: ./celp -i inputfile.wav -o outputfile
    	\item \emph{Synthesis}: ./celp -c outputfile.chan -o outputsynth
    	\item \emph{Copy spd (speech data) file to raw file}: cp outputsynth.spd outputsynth.raw
    	\item \emph{Convert to wav file}: sox -r 8000 -b 16 -c 1 -e signed-integer outputsynth.raw outputsynth.wav
    	\item \emph{Playing the file}: padsp play outputsynth.wav
  	\end{enumerate}
  \item LD-CELP commands
	\begin{enumerate}
    	\item \emph{Analysis}: ./ccelp inputfile.wav encoderout.out
    	\item \emph{Synthesis}: ./dcelp encoderout.out outputsynth.raw
 	\end{enumerate}
  \item MELP commands
	\begin{enumerate}
    	\item \emph{Analysis}: ./melp -a -i inputfile.wav -o encoderout.out
  	\item \emph{Synthesis}: ./melp -s -i encoderout.out -o outputsynth.raw
  	\end{enumerate}
\end{enumerate}
\indent In LD-CELP and MELP, conversion of headerless raw format to wav file format and playing of the output file are performed in the same manner as that of CELP.
%-----------------------------------------------------------------------------------------------------------------------
\subsection{Waveform analysis}
Waveform analysis was performed using Praat, a tool used for phonetic analysis of speech. A standard speech sample \emph{source.wav} was used for waveform analysis. Time domain representation of the speech files as well as pitch and intensity waveforms were plotted using the Praat Objects and Picture windows. Spectrogram analysis was done with the help of Adobe Audition software.

%-----------------------------------------------------------------------------------------------------------------------

\subsection{Subjective Testing: Mean Opinion Score}
Evaluation of the perceptual quality was done using the Mean Opinion Score (MOS) test. Due to time constraints, informal testing of the codecs was conducted with 10 evaluators. Speech samples recorded in the English language were given as input to the vocoders. Three samples were recorded in a quiet environment, while two speech samples were recorded with loud background music. Logitech h110 stereo headsets were used for voice recording and playback. The evaluators were given an initial training on the MOS test, and their scores were recorded. MOS scores and their interpretations are given below \cite{voip}. 
\begin{center}
\begin{tabular}{|c|c|}
\hline
\textbf{MOS} & \textbf{Quality} \\
\hline
	5  & Excellent \\ 
	4 & Good  \\ 
	3 & Fair \\ 
	2  & Poor  \\ 
	1  & Bad  \\ \hline
\end{tabular}
\label{tab:mos}
\end{center}
A MOS score of 4 or 5 indicates toll quality speech while a score of 1 or 2 indicates synthetic speech.
\begin{figure*}[!t]
\centerline{\subfloat[Comparison of pitch contours]{\includegraphics[scale=0.14]{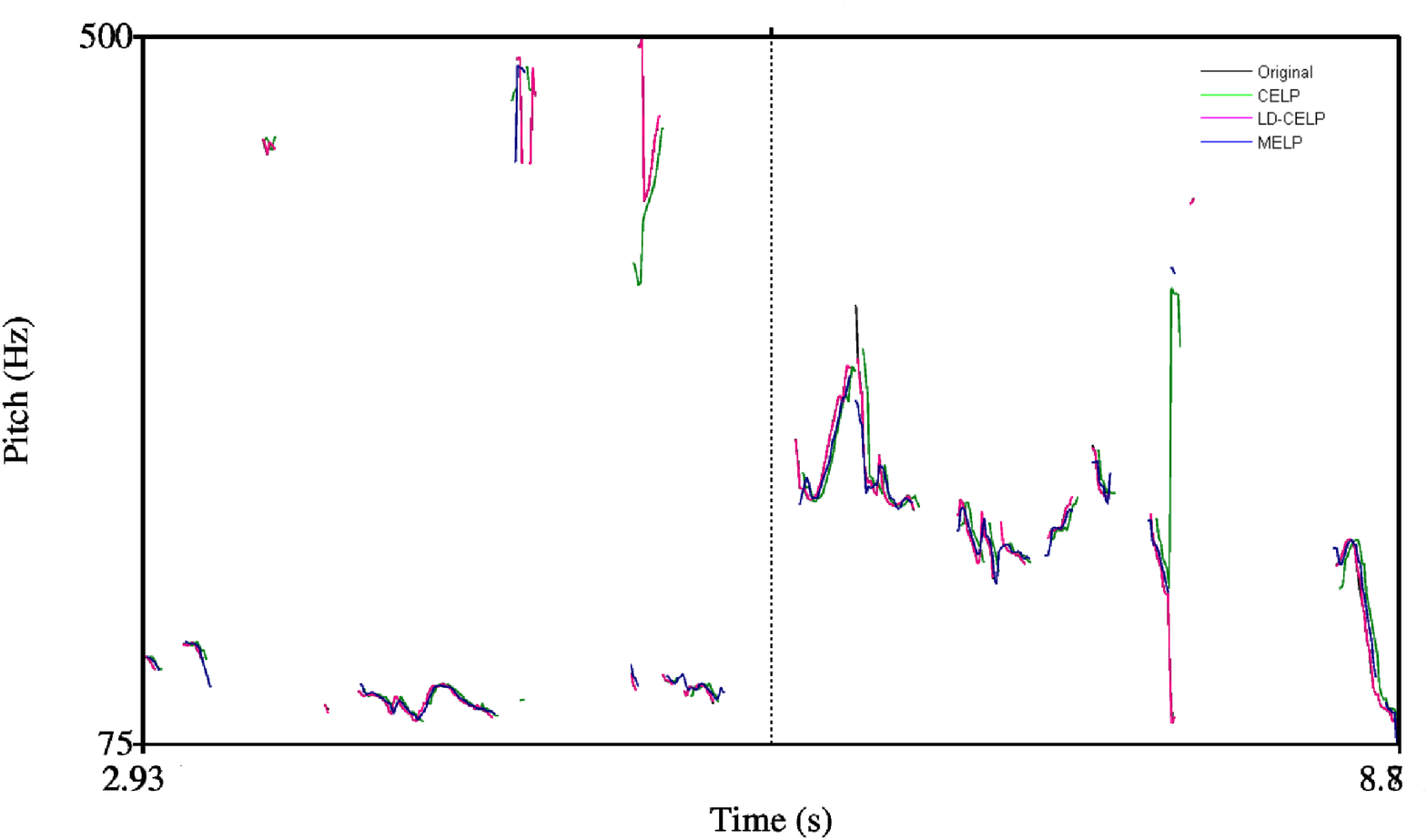}%
\label{fig:pitch}}
\hfil
\subfloat[Comparison of intensity contours]{\includegraphics[scale=0.143]{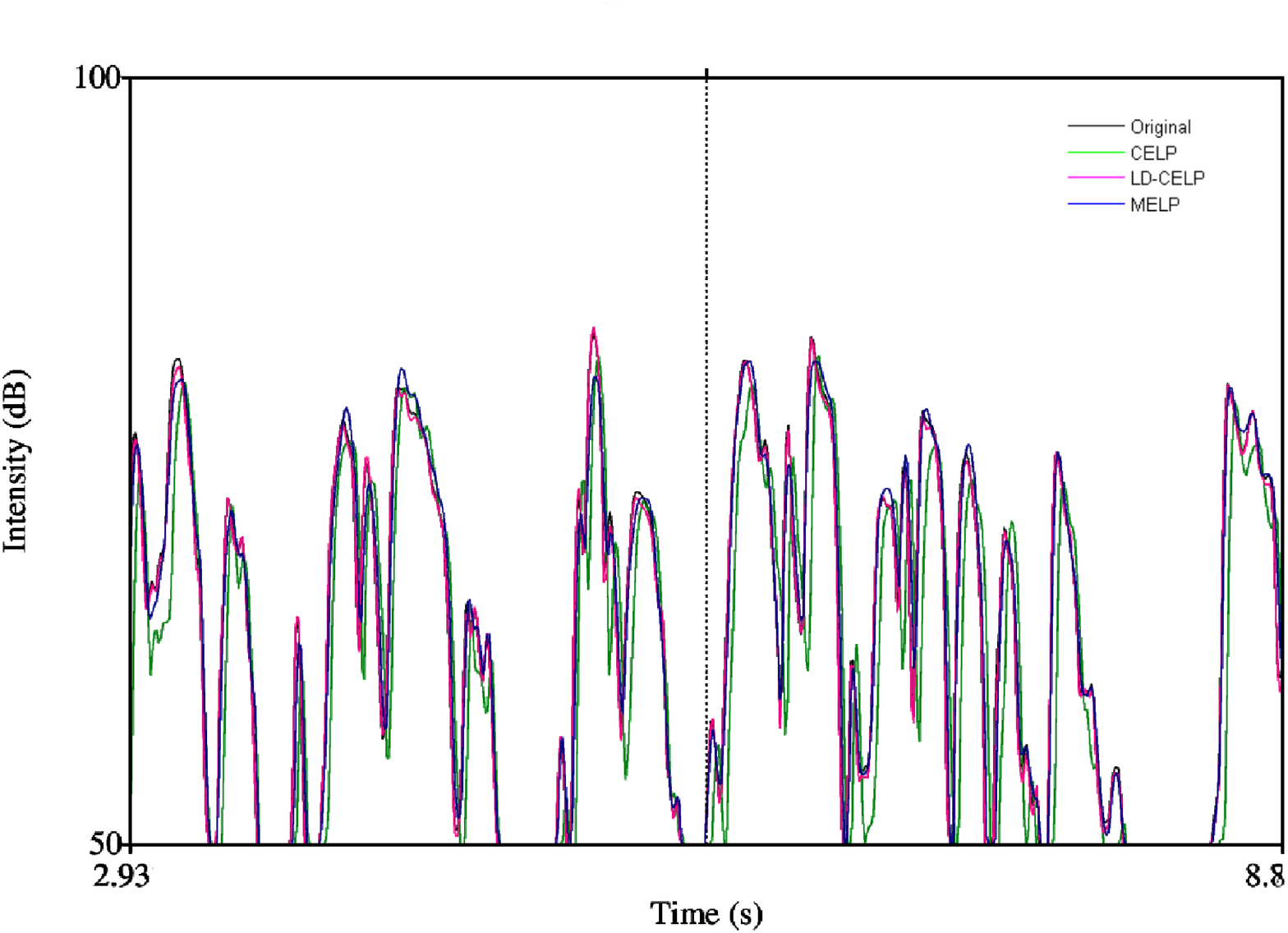}%
\label{fig:intensity}}}
\caption{Pitch and intensity waveforms}
\label{fig:p_int}
\end{figure*}

%-----------------------------------------------------------------------------------------------------------------------

\section{Results and Discussion}
\subsection{Waveform analysis}
The results of waveform analysis using Praat software are shown in Figs.~\ref{fig:time} to~\ref{fig:p_int}. Fig.~\ref{fig:time} shows the time domain representation of the original and synthesized speech files of CELP, LD-CELP and MELP coders. Fig.~\ref{fig:spectrogram} depicts the spectrograms of the original and synthesized speech waveforms. Fig.~\ref{fig:p_int} shows the pitch and intensity contours. 
\begin{table}[!t]
\renewcommand{\arraystretch}{1.3}
% if using array.sty, it might be a good idea to tweak the value of
% \extrarowheight as needed to properly center the text within the cells
\caption{Input speech file details}
\label{tab:in_speech_files}
\centering
%% Some packages, such as MDW tools, offer better commands for making tables
%% than the plain LaTeX2e tabular which is used here.
\begin{tabular}{ccc}
\hline
Filename & File Size (kB) & Duration (s)\\
\hline
\hline
	male\_eng.wav & 170 & 10.65 \\
	female\_eng.wav & 176 & 10.98 \\ 	
	male\_fem\_conversation.wav & 319   & 19.91 \\ 
	male\_noisy\_eng.wav & 447 & 27.93 \\ 
	female\_noisy\_eng.wav & 856 & 53.49 \\ \hline
\end{tabular}
\end{table}

\begin{table}[!t]
\renewcommand{\arraystretch}{1.3}
% if using array.sty, it might be a good idea to tweak the value of
% \extrarowheight as needed to properly center the text within the cells
\caption{MOS score for vocoders}
\label{tab:MOS}
\centering
%% Some packages, such as MDW tools, offer better commands for making tables
%% than the plain LaTeX2e tabular which is used here.
\begin{tabular}{cccc}
\hline
Filename & CELP & LD-CELP & MELP\\
\hline
\hline
	male\_eng.wav & 3.10 & 4.06 & 2.82 \\
	female\_eng.wav & 3.24 &  4.02 & 2.76 \\ 	
	male\_fem\_conversation.wav & 3.10 & 3.76    & 3.12 \\ 
	male\_noisy\_eng.wav & 3.00 & 4.58 & 3.24 \\ 
	female\_noisy\_eng.wav & 2.52 & 4.26 &  1.72 \\ \hline
	\textbf{Average MOS}	& \textbf{2.992} & \textbf{4.136} &  \textbf{2.732}\\ \hline
\end{tabular}
\end{table}

From the time domain waveforms, it can be concluded that the overall shape of the original wave has been preserved. However peaks have been clipped at certain portions which result in decrease in clarity of the speech output. The spectrograms show the frequency content of the speech waveforms as a function of time. The pitch and intensity graphs also show slight variations in the output of the vocoders. Reliable estimates of the perceptual quality can be made only by conducting subjective tests using human listeners. 

%-----------------------------------------------------------------------------------------------------------------------
\subsection{Subjective Testing: Mean Opinion Score}
All recorded input speech files were sampled at a rate of 8 kHz. The details of speech input files are shown in Table~\ref{tab:in_speech_files}. The average input speech file size was 393.6kB and average duration 24.59s. The MOS scores corresponding to each input speech file and the average MOS score obtained for each vocoder are shown in Table~\ref{tab:MOS}.
\newline \indent The MOS scores in Table~\ref{tab:MOS} show that LD-CELP has the highest perceptual quality (toll quality) among the three vocoders. The perceptual quality of CELP and MELP vocoders is rated less, with CELP scoring slightly higher than MELP.

%-----------------------------------------------------------------------------------------------------------------------
\section{Conclusion}

The results of the comparison indicate that a choice can be made only based on the application of the vocoder. In applications where the focus is on low delay and high perceptual quality, as in two-way communication systems, the LD-CELP algorithm at 16 kb/s is the ideal candidate. In areas where low bit rate is essential, MELP is the best candidate because it can work at bit rates as low as 2.4 kb/s and gives intelligible output.   When both low bit rate and good quality are required, the CELP coder at 4.8 kb/s seems to be the most suitable coder. In this study, the number of evaluators for the MOS test was limited to 10 due to time constraints. For more accurate results, the number of evaluators needs to be increased. 

\section*{Acknowledgment}

The authors would like to thank the students and faculty of the department for providing speech samples and for their participation in the Mean Opinion Score testing. Special thanks goes to Karthika Balan for her valuable help in the MOS testing process.

%-----------------------------------------------------------------------------------------------------------------------

%\IEEEtriggeratref{1}


%\bibliographystyle{IEEEtran}

\end{document}